\documentclass[onecolumn,secnumarabic,amssymb, amsmath, 12pt,
baselineskip=30pt,
nofootinbib,tightenlines,  nobibnotes, aps, prl,epsfig]{revtex4}
\usepackage{graphicx}
\usepackage{dcolumn}
\usepackage{bm}
\begin{document}
\title{The ratio of  the charm structure functions $F_{k}^{c}(k=2,L)$ at low-$x$ in DIS with respect to the expansion method}

\author{B.Rezaei}%
 \email{brezaei@razi.ac.ir }
\author{G.R.Boroun }
\altaffiliation{grboroun@gmail.com; boroun@razi.ac.ir}
\affiliation{ Physics Department, Razi University, Kermanshah
67149, Iran}
\date{\today}
\begin{abstract}
We study the expansion method to the gluon distribution function
 at low $x$ values and calculate the charm structure functions in LO and NLO
 analysis. Our results provide compact formula for the ratio $R^{c}=\frac{F_{L}^{c}}{F^{c}_{2}}$  that is
approximately independent of $x$ and the details of the parton
distribution function at low $x$ values. This ratio could be good
probe of the charm structure function $F^{c}_{2}$ in the proton
from the reduced charm cross sections at DESY HERA. These results
show that the charm structure functions obtained are in agreement
with HERA experimental data and other theoretical models.
\end{abstract}
 \pacs{13.60.Hb; 12.38.Bx}
\keywords{Charm Structure Function; Gluon Distribution;
Quantum Chromodynamics; Small-$x$} 
\maketitle
\subsection{Introduction}

The low- $x$ regime of the quantum Cheromodynamic (QCD) has been
intensely investigated in recent years for consideration of the
heavy quarks [1-5]. Of course the notation of the intrinsic charm
content of the proton has been introduced over 30 years ago in
Ref.[6]. The study of production mechanisms of heavy quarks
provides us with new tests of QCD. As in perturbative QCD (pQCD),
physical quantities can be expanded into the strong coupling
constant $\alpha_{s}(\mu^{2})$. Extensive the $\mu$ scale to the
large values establish the theoretical analysis as can be
described with hard processes. In the case of heavy quark
production, we can have condition that the heavy quarks produced
from the boson- gluon fusion (BGF) according to Fig.1. That is, in
PQCD calculations the production of heavy quarks at HERA proceeds
dominantly via the direct BGF where the photon interacts with
a gluon from the proton by the exchange of a heavy quark pair.\\

In this processes all quark flavours lighter than charm are
treated as massless with massive charm being produced dynamically
in BGF. Charm production contributes to the total deep inelastic
scattering (DIS) cross section by at most $30\%$ at HERA [7]. In
the recent measurements of HERA [8], the charm contribution to the
structure function at small $x$ is a large fraction of the total.
This behavior is directly related to the growth of the gluon
density at small $x$, as gluons couple only through the strong
interaction. Consequently the gluons are not directly probed in
DIS, only contributing indirectly via the
 $g{\rightarrow}q\bar{q}$
transition. This involves the computation of the BGF process
$\gamma^{\star}g{\rightarrow}c\bar{c}$. This process can be
created when the squared invariant mass of the hadronic final
state is $W^{2}{\geq}4m_{c}^{2}$.\\
 In this paper we apply the
expansion of the gluon distribution at an arbitrary point to the
charm structure functions in deep inelastic scattering. Then we
present the ratio of the charm structure functions, that is
independent of the gluon distribution and its useful to extract
the charm structure function from the reduced charm cross section
experimental data.\\

\subsection{Charm components of the structure functions}

In deeply inelastic electron- proton scattering, the heavy- quark
contribution to heavy flavor is according to this reaction
\begin{equation}
e(l_{1})+P(p){\rightarrow}e(l_{2})+Q(p_{1})\overline{Q}(p_{2})+X.
\end{equation}
Here, neglecting the contribution of Z- boson exchange and
omitting charged- current interactions. The deeply inelastic
electroproduction cross section for the heavy quark-antiquark in
the final sate can be written as
\begin{eqnarray}
\frac{d^{2}\sigma^{c\overline{c}}}{dxdQ^{2}}&=&\frac{2{\pi}\alpha^{2}}{xQ^{4}}(1+(1-y)^{2})F_{2}(x,Q^{2},m_{c}^{2})\\\nonumber
&&\times[1-\frac{y^{2}}{1+(1-y)^{2}}R^{c}],
\end{eqnarray}
where $R^{c}$ denotes the ratio of the charm structure functions
and the kinematic variables are defined by $x=\frac{Q^{2}}{2p.q},
y=\frac{p.q}{p.l}$ and $Q^{2}=-q^{2}$.\\

 The deeply inelastic
charm structure functions ($F_{k}(x,Q^{2},m_{c}^{2})$ for $k=2,L$)
in the cross section (2) is given by [9]
\begin{eqnarray}
F_{k}^{c}(x,Q^{2},m^{2}_{c})=2xe_{c}^{2}\frac{\alpha_{s}(\mu^{2})}{2\pi}\int_{ax}^{1}\frac{dy}{y}C_{g,k}
(\frac{x}{y},\zeta)g(y,\mu^{2}),
\end{eqnarray}
where $a=1+4\zeta(\zeta{\equiv}\frac{m_{c}^{2}}{Q^{2}})$,
$g(x,\mu^{2})$ is the gluon density and the mass factorization
scale $\mu$, which has been put equal to the renormalization
scale, is assumed to be either $\mu^{2}=4m_{c}^{2}$ or
$\mu^{2}=4m_{c}^{2}+Q^{2}$. Here $C^{c}_{g,k}$ is the charm
coefficient function in LO and NLO analysis as
\begin{eqnarray}
C_{k,g}(z,\zeta)&{\rightarrow}&C^{0}_{k,g}(z,\zeta)+a_{s}(\mu^{2})[C_{k,g}^{1}(z,\zeta)\\\nonumber
&&+\overline{C}_{k,g}^{1}(z,\zeta)ln\frac{\mu^{2}}{m_{c}^{2}}],
\end{eqnarray}
where $a_{s}(\mu^{2})=\frac{\alpha_{s}(\mu^{2})}{4\pi}$ and in the
NLO analysis
\begin{eqnarray}
\alpha_{s}(\mu^{2})=\frac{4{\pi}}{\beta_{0}ln(\mu^{2}/\Lambda^{2})}
-\frac{4\pi\beta_{1}}{\beta_{0}^{3}}\frac{lnln(\mu^{2}/\Lambda^{2})}{ln(\mu^{2}/\Lambda^{2})}
\end{eqnarray}
with $\beta_{0}=11-\frac{2}{3}n_{f},
\beta_{1}=102-\frac{38}{3}n_{f} $ ($n_{f}$ is the number of active
flavours).\\

In the LO analysis, the coefficient functions BGF can be found
[9], as
\begin{eqnarray}
C^{0}_{g,2}(z,\zeta)&=&\frac{1}{2}([z^{2}+(1-z)^{2}+4z\zeta(1-3z)-8{\zeta^{2}}z^{2}]\nonumber\\
&&{\times}ln\frac{1+\beta}{1-\beta}+{\beta}[-1+8z(1-z)\nonumber\\
&&-4z{\zeta}(1-z)]),
\end{eqnarray}
and
\begin{eqnarray}
C^{0}_{g,L}(z,\zeta)=-4z^{2}{\zeta}ln\frac{1+\beta}{1-\beta}+2{\beta}z(1-z),
\end{eqnarray}
where $\beta^{2}=1-\frac{4z\zeta}{1-z}$.\\
 At NLO,
$O(\alpha_{em}\alpha_{s}^{2})$, the contribution of the photon-
gluon component is usually presented in terms of the coefficient
functions $C_{k,g}^{1}, \overline{C}_{k,g}^{1}$. Using the fact
that  the virtual photon- quark(antiquark) fusion subprocess can
be neglected, because their contributions to the heavy-quark
leptoproduction vanish at LO and are small at NLO [1,10].  In a
wide kinematic range, the contributions to the charm structure
functions in NLO are not positive due to mass factorization and
are less than $\%10$. Therefore the charm structure functions are
dependence to the gluonic observable in LO and NLO. The NLO
coefficient functions are only available as computer codes[9,10].
But in the high- energy regime ($\zeta<<1$) we can used the
compact form of these coefficients according to
the Refs.[11,12].\\

\subsection{The method}

Now we want to calculate the charm structure functions by using
the expansion method for the gluon distribution function. As can
be seen, the dominant contribution to the charm structure
functions comes from the gluon density at small $x$, regardless of
the exact shape of the gluon distribution. Substitute
$y=\frac{x}{1-z}$ in Eq.3 to obtain the more useful form, as
\begin{eqnarray}
F_{k}^{c}(x,Q^{2},m^{2}_{c})&=&2e_{c}^{2}\frac{\alpha_{s}(\mu^{2})}{2\pi}\int_{1-\frac{1}{a}}^{1-x}dzC_{g,k}^{c}
(1-z,\zeta)\nonumber\\
&& {\times}G(\frac{x}{1-z},\mu^{2}),
\end{eqnarray}
here $G(x)=xg(x)$ is the gluon distribution function. The argument
$\frac{x}{1-z}$ of the gluon distribution in Eq.8 can be expanded
at an arbitrary point $z=\alpha$ as
\begin{equation}
\frac{x}{1-z}|_{z=\alpha}=\frac{x}{1-\alpha}\sum_{k=1}^{\infty}[1+\frac{(z-\alpha)^{k}}{(1-\alpha)^{k}}].
\end{equation}
The above series is convergent for $|z-\alpha|<1$. Using this
expression we can rewrite and expanding the gluon distribution
$G(\frac{x}{1-z})$  as
\begin{eqnarray}
G(\frac{x}{1-z})&=&G(\frac{x}{1-\alpha})\\\nonumber
&&+\frac{x}{1-\alpha}(z-\alpha)\frac{{\partial}G(\frac{x}{1-a})}{{\partial}x}+O(z-\alpha)^{2}.
\end{eqnarray}
Retaining terms only up to the first derivative in the expansion
and doing the integration, we obtain our master formula as
\begin{eqnarray}
F_{k}^{c}(x,Q^{2},m^{2}_{c})=2e_{c}^{2}\frac{\alpha_{s}(\mu^{2})}{2\pi}A_{k}(x)\nonumber\\
{\times}G(\frac{x}{1-\alpha}(1-\alpha+\frac{B_{k}(x)}{A_{k}(x)})),
\end{eqnarray}
where
\begin{eqnarray}
A_{k}(x)=\int_{1-\frac{1}{a}}^{1-x}C_{g,k}^{c} (1-z,\zeta)dz,
\end{eqnarray}
and
\begin{eqnarray}
B_{k}(x)=\int_{1-\frac{1}{a}}^{1-x}(z-\alpha)C_{g,k}^{c}(1-z,\zeta)dz.
\end{eqnarray}
where $C_{g,k}^{c}$ is defined at Eq.4 in LO and NLO analysis and
also $\alpha$ has an arbitrary value $0{\leq}\alpha{<}1$. Eq.10
can be rewritten as
\begin{eqnarray}
F_{k}^{c}(x,Q^{2},m^{2}_{c})=2e_{c}^{2}\frac{\alpha_{s}(\mu^{2})}{2\pi}\eta_{k}
G(\frac{x}{1-\alpha}({\beta_{k}}-\alpha ),\mu^{2}).
\end{eqnarray}
This result shows that the charm structure functions
$F_{k}^{c}(x,Q^{2})$ at $x$ are calculated using the gluon
distribution at $\frac{x}{1-\alpha}(\beta_{k}-\alpha)$. Therefore,
the gluon distribution at $\frac{x}{1-\alpha}(\beta_{k}-\alpha)$
can be simply extracted the charm structure functions ($F_{2}^{c}$
and $F_{L}^{c}$) in the low $x$ values according to the
coefficients at the limit $x{\rightarrow}0$, in Table 1. Moreover,
there is a directly relation between the charm structure functions
and gluon distribution via the well known Bethe- Heitler process
$\gamma^{\star}g{\rightarrow}c\bar{c}$.\\

Now, we defining the ratio of the charm structure functions and
using Eq.14, we obtain the following equation
\begin{eqnarray}
R^{c}=\frac{\eta_{L}}{\eta_{2}}\frac{G(\frac{x}{1-\alpha}(\beta_{L}-\alpha)}{G(\frac{x}{1-\alpha}(\beta_{2}-\alpha)}.
\end{eqnarray}
We observe that the right-hand side of this ratio is independent
of $x$ and independent of the gluon distribution input according
to the coefficients in Table 1. As in the low $x$ range we have
\begin{eqnarray}
R^{c}{\approx}\frac{\eta_{L}}{\eta_{2}},
\end{eqnarray}
 which is very useful to extract the charm structure function $F_{2}^{c}(x,Q^{2})$ from
 measurements of the doubly differential cross section of inclusive
  deep inelastic scattering at DESY HERA, independent of the gluon
distribution function. Therefore, we can determine the charm
structure function into the reduced cross section from the double-
differential charm cross section as
\begin{eqnarray}
F_{2}(x,Q^{2},m_{c}^{2})=\widetilde{\sigma}^{c\overline{c}}(x,Q^{2})/[1-\frac{y^{2}}{1+(1-y)^{2}}R^{c}].
\end{eqnarray}
Where $R^{c}$ is defined at Eq.16 and Table 1 and
$\widetilde{\sigma}^{c\overline{c}}$ is taking from Ref.13, also
the error bars in our determination  can be examined by the
following expression (Table 1)
\begin{eqnarray}
\delta_{F_{2}^{c\overline{c}}}=F_{2}^{c\overline{c}}[\frac{\delta_{\widetilde{\sigma}^{c\overline{c}}}}
{\widetilde{\sigma}^{c\overline{c}}}+\frac{\frac{y^{2}}{1+(1-y)^{2}}\delta_{R^{c}}}{1-\frac{y^{2}}{1+(1-y)^{2}}R^{c}}].
\end{eqnarray}


\subsection{Results and Discussion}

For the calculation of the charm structure functions
($F_{2}^{c\overline{c}}$ and $F_{L}^{c\overline{c}}$), we choose
 $\Lambda=0.224 GeV$, $m_{c}=1.5GeV$ and known that the dominant uncertainty in the
QCD calculations arises from the uncertainty in the charm quark
mass. Since the contribution of the longitudinal charm structure
function to the DIS charm cross section (i.e., Eq.(2)) is
proportional to $y^2$, so that the $F_{2}^{c\overline{c}}$ term
dominates at $ y\le{0.08 }$ and the relation $\widetilde{\sigma
}^{c\overline{c}}=F_{2}^{c\overline{c}}$ holds to a very good
approximation. Thus the contribution of the second term of the
right hand Eq.(2) can be sizeable only at $y>{0.08}$. Therefore,
for $y>{0.08}$, the ratio of the charm structure functions is very
useful. In Fig.2 we observe that this ratio is according to
results Refs.4 and 11 at low $x$. Also at NLO analysis its
decrease as $Q^{2}$ increases and this is familiar from the
Callan- Gross ratio. As we can see in this figure, this ratio has
value $0.1<R^{c}<0.2$ in a wide region of $Q^{2}$.\\
We now extract $F_{2}^{c\overline{c}} $ from the H1 measurements
of the reduced charm cross section [13] in Eq.17 with respect to
Eq.16 for $Q^{2}{\geq}8.5 GeV^{2}$. Our NLO results for the charm
structure function are presented in Table 2, where they are
compared with the experimental values from H1 data and they are
comparable with the HVQDIS and CASCADE programs [14,15] as we can
see at Table 11 in Ref.13(arXiv:1106.1028v1 [hep-ex] 6 Jun 2011).
The error bars in Table 2 are according to the theoretical
uncertainty related to the freedom in the choice of the
renormalization scales in the ratio of the charm structure
function and also the experimental total errors related to the
results in Ref.13 according to the Eq.18. A comparison between our
obtained values for the charm structure function and the existing
data, indication the fact that the ratio $R^{c}$ can be determined
with reasonable precision at
any $y$ value.\\

In order to test the validity and correctness of our obtained
charm structure functions with respect to the gluon distribution
function (Eq.14), we obtained the charm structure functions into
 the gluon distribution input, which is usually taken from NLOGRV [9] or Block [16] parameterizations.
As the gluon distribution input is dependent to a point of
expansion $\alpha$. In order to estimate the theoretical
uncertainty resulting from this, we choose $\alpha=0$ and
$\alpha=0.8$ in the renormalization scale $\mu^{2}=4mc^{2}+Q^{2}$.
In Figs.3-6, we observe that the theoretical uncertainty related
to the freedom in the choice of $\alpha$ is very small at the
renormalization scales. As can be seen in Figs.3-4, the better
choice of the expansion point for the charm structure function
$F_{2}^{c}$ is at the point $\alpha{\simeq}0.5$, as this point is
favoured according to the current data. This means that in this
kinematical  region the longitudinal momentum of the gluon $x_{g}$
is more than three times the value of the longitudinal momentum of
the probed charm quark- anitquark in BGF process. We compared our
results for the charm structure function to the DL model [17-19],
H1 data [13] and color dipole model [20]. In Figs.5-6, the better
choice of the expansion point for the longitudinal charm structure
function $F_{L}^{c}$  is $\alpha{\geq}0.8$, as compared only to
the color dipole model [20]. As can be seen in these figures, the
increase of our results for the
 charm structure functions $F^{c}_{k}(x,Q^{2})$ towards low
 $x$ are consistent and comparable with the experimental data and theoretical
 models.\\

\subsection{Conclusion}
In summary, we have used the expansion method for the low $x$
gluon distribution and derived a compact formula for the  ratio
$R^{c}=\frac{F_{L}^{c\overline{c}}}{F_{2}^{c\overline{c}}}$ of the
charm structure functions at NLO analysis. We observed that the
this ratio is independent of $x$ and independent of the parton
distribution function input, and also its useful to extract the
charm structure function from the reduced charm cross section.
Based upon of the reduced charm cross section in the low $x$
region, an approximate method for the calculation of the charm
structure function $F_{2}^{c\overline{c}}$ is presented. Careful
investigation of our results show a good agreement with the recent
published charm structure functions $F_{2}^{c\overline{c}}$ and
other theoretical models within errors from the expansion point and the renormalization scales.\\

\newpage
\textbf{References}\\
1. A.Vogt, arXiv:hep-ph:9601352v2(1996).\\
2. H.L.Lai and W.K.Tung, Z.Phys.C\textbf{74},463(1997).\\
3. A.Donnachie and P.V.Landshoff, Phys.Lett.B\textbf{470},243(1999).\\
4. N.Ya.Ivanov, Nucl.Phys.B\textbf{814}, 142(2009);  N.Ya.Ivanov
and B.A.Kniehl, Eur.Phys.J.C\textbf{59}, 647(2009).\\
5. F.Carvalho, et.al., Phys.Rev.C\textbf{79}, 035211(2009).\\
6. S.J.Brodsky, P.Hoyer, C.Peterson and
N.Sakai,Phys.Lett.B\textbf{93}, 451(1980); S.J.Brodsky, C.Peterson
and N.Sakai, Phys.Rev.D\textbf{23}, 2745(1981).\\
7. K.Lipta, PoS(EPS-HEP)313,(2009).\\
8. C. Adloff et al. [H1 Collaboration], Z. Phys. C\textbf{72}, 593
(1996); J. Breitweg et al. [ZEUS Collaboration], Phys. Lett.
B\textbf{407}, 402 (1997); C. Adloff et al. [H1 Collaboration],
Phys. Lett. B\textbf{528}, 199 (2002); S. Aid et. al., [H1
Collaboration], Z. Phys. C\textbf{72}, 539 (1996); J. Breitweg et.
al., [ZEUS Collaboration], Eur. Phys. J. C\textbf{12}, 35 (2000);
S. Chekanov et. al., [ZEUS Collaboration], Phys. Rev.
D\textbf{69}, 012004 (2004); Aktas et al. [H1 Collaboration], Eur.
Phys.J. C\textbf{45}, 23 (2006); F.D. Aaron et al. [H1
Collaboration],Eur.Phys.J.C\textbf{65},89(2010).
\\
9. M.Gluk, E.Reya and A.Vogt, Z.Phys.C\textbf{67}, 433(1995); Eur.Phys.J.C\textbf{5}, 461(1998). \\
10. E.Laenen, S.Riemersma, J.Smith and W.L. van Neerven,
Nucl.Phys.B \textbf{392}, 162(1993).\\
11. A.~Y.~Illarionov, B.~A.~Kniehl and A.~V.~Kotikov, Phys.\
Lett.\  B {\bf 663}, 66 (2008).\\
12. S. Catani, M. Ciafaloni and F. Hautmann, Preprint
CERN-Th.6398/92, in Proceeding of the Workshop on Physics at HERA
(Hamburg, 1991), Vol. 2., p. 690; S. Catani and F. Hautmann, Nucl.
Phys. B \textbf{427}, 475(1994); S. Riemersma, J. Smith and W. L.
van Neerven, Phys. Lett. B \textbf{347}, 143(1995).\\
 13. F.D.Aaron, et.al,. H1 Collab.,
 Phys.Lett.b\textbf{665},139(2008); Eur.Phys.J.C\textbf{71}, 1509(2011); Eur.Phys.J.C\textbf{71}, 1579(2011); arXiv:1106.1028v1 [hep-ex] 6 Jun 2011;
  arXiv:0911.3989v1  [hep-ex] 20 Nov 2009.\\
14. H. Jung, CASCADE V2.0, Comp. Phys. Commun. \textbf{143}, 100(2002).\\
15. B.W. Harris and J. Smith, Nucl. Phys. B\textbf{452},
109(1995); Phys. Rev. D\textbf{57}, 2806(1998).\\
16. M.M.Block, L.Durand and D.W.Mckay, Phys.Rev.D\textbf{77},
 094003(2008).\\
17. A.Donnachie and P.V.Landshoff, Z.Phys.C \textbf{61},
139(1994); Phys.Lett.B \textbf{518}, 63(2001); Phys.Lett.B
\textbf{533}, 277(2002); Phys.Lett.B \textbf{470}, 243(1999);
Phys.Lett.B \textbf{550}, 160(2002).\\
18. R.D.Ball and P.V.landshoff, J.Phys.G\textbf{26}, 672(2000).\\
19. P.V.landshoff, arXiv:hep-ph/0203084 (2002).\\
20. N.N.Nikolaev and V.R.Zoller, Phys.Lett. B\textbf{509},
283(2001).\\
\newpage{
\subsection{Figure captions }
Fig.1: The photon- gluon fusion.\\

   Fig.2: The ratio $R^{c}$ evaluated as function of $Q^{2}$ at
NLO analysis from Eq.16. The error bars are the theoretical
uncertainty using
 the renormalization scales $\mu^{2}=4m_{c}^{2}$ and $\mu^{2}=4m_{c}^{2}+Q^{2}$.\\

 Fig.3: The charm structure function ($F_{2}^{c\overline{c}} $)
obtained at $Q^{2}=20 GeV^{2}$ with respect to the input gluon
distribution NLO-GRV parametermization [9] (Solid line according
to the expanding point $\alpha=0$ and Dash-Dot line according to
the expanding point $\alpha=0.8$) compared with DL fit[17-19] (Dot
line), color dipole model [20] (Dash line) and H1 data [13]
(square) that accompanied with total errors
 at
 the renormalization scale $\mu^{2}=4m_{c}^{2}+Q^{2}$.\\

 Fig.4: The charm structure function ($F_{2}^{c\overline{c}} $)
obtained at $Q^{2}=20 GeV^{2}$ with respect to the input gluon
distribution Block fit [16] (Solid line according to the expanding
point $\alpha=0$ and Dash-Dot line according to the expanding
point $\alpha=0.8$) compared with DL fit[17-19] (Dot line), color
dipole model [20] (Dash line) and H1 data [13] (square) that
accompanied with total errors at
 the renormalization scale $\mu^{2}=4m_{c}^{2}+Q^{2}$.\\

 Fig.5: The longitudinal charm structure function
($F_{L}^{c\overline{c}} $) obtained at $Q^{2}=20 GeV^{2}$ with
respect to the input gluon distribution  NLO-GRV parametermization
[9] (Solid line according to the expanding point $\alpha=0$ and
Dash-Dot line according to the expanding point $\alpha=0.8$)
compared with the color dipole model [20] (Dash line) at
 the renormalization scale $\mu^{2}=4m_{c}^{2}+Q^{2}$.\\

 Fig.6: The longitudinal charm structure function
($F_{L}^{c\overline{c}} $) obtained at $Q^{2}=20 GeV^{2}$ with
respect to the input gluon distribution Block fit [16] (Solid line
according to the expanding point $\alpha=0$ and Dash-Dot line
according to the expanding point $\alpha=0.8$) compared with the
color dipole model [20] (Dash line) at
 the renormalization scale $\mu^{2}=4m_{c}^{2}+Q^{2}$.\\}

\begin{figure}
\includegraphics{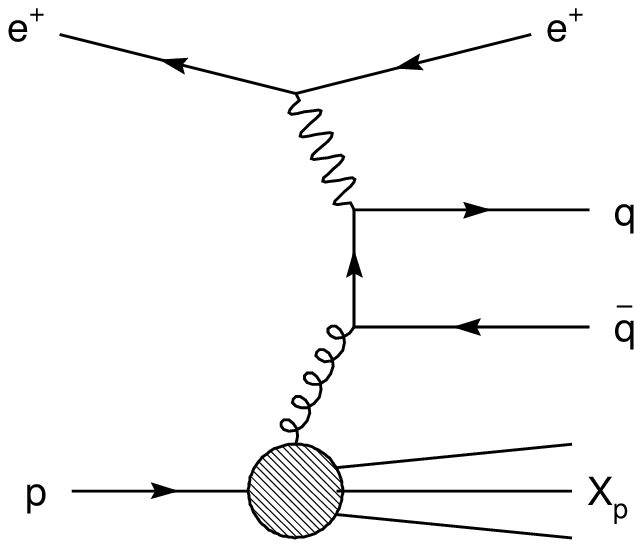}
\caption{} \label{Fig1}
\end{figure}
\begin{figure}
\includegraphics{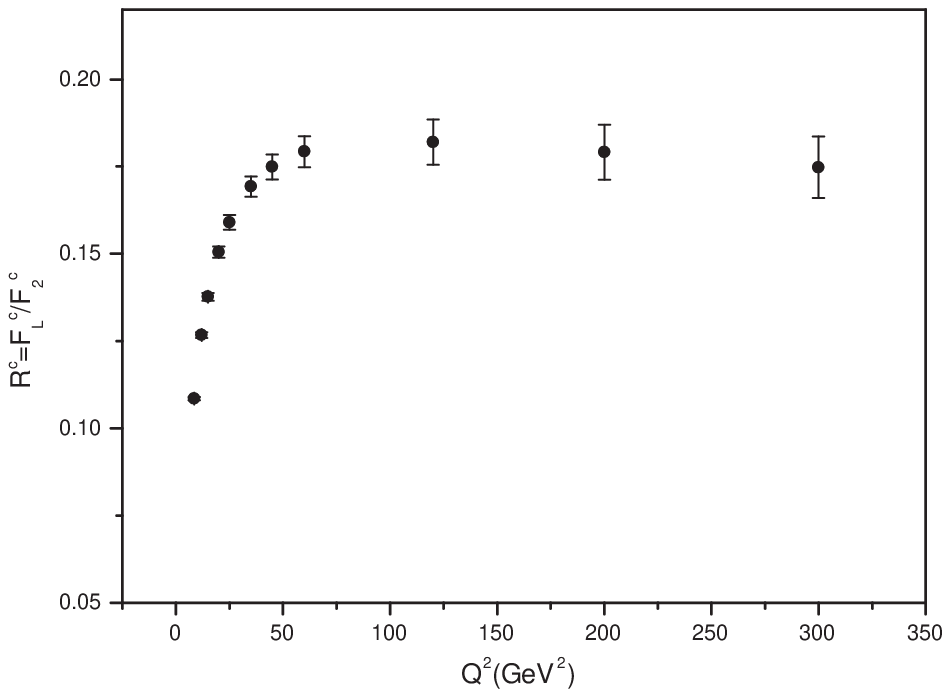}
\caption{  }\label{Fig2}
\end{figure}
\begin{table}
\centering \caption{The constant values in this analysis at
$Q^{2}$ values in the limit $x{\rightarrow}0$.
}\label{table:table1}
\begin{minipage}{\linewidth}
\renewcommand{\thefootnote}{\thempfootnote}
\centering
\begin{tabular}{|l||c|c||c|c||c|c||c|c||c|c|} \hline\noalign{\smallskip} $Q^{2}(GeV^{2})$ & $ \eta_{2}$ &
$ \delta_{\eta_{2}}$ & $ \beta_{2}$ & $ \delta_{\beta_{2}}$ &
$\eta_{L}$ &$ \delta_{\eta_{L}}$ & $\beta_{L}$ &
 $ \delta_{\beta_{L}}$ & $R^{c}$ & $\delta_{R^{c}}$ \\
\hline\noalign{\smallskip}
8.5& 0.4645 & 1.95E-3 & 1.8393 & 2E-4 & 0.0504 & 4E-4 &1.7853 &1E-4   &0.1085  &4.5E-4 \\
12 & 0.5763 & 3.7E-3 & 1.8083 & 3.5E-4& 0.0730 & 9E-4 &1.7453 &1.5E-4 &0.1267 &8E-4 \\
15 & 0.6546 & 5.45E-3 & 1.7883 & 5E-4 & 0.0901 & 1.45E-3 &1.7200&1E-4&0.1377&1.1E-3 \\
20 & 0.7611 & 8.45E-3 & 1.7632 & 7E-4 & 0.1145 & 2.45E-3 &1.6883&2E-4&0.1505&1.6E-3 \\
25 & 0.8469 & 0.0114 & 1.7447 & 1E-3& 0.1347 & 3.6E-3&1.6654&2.5E-4&0.1590&2.1E-3 \\
35 & 0.9795 & 0.0173 & 1.7190 & 1.35E-3& 0.1659 & 5.8E-3&1.6343&3E-4&0.1693&2.9E-3 \\
45 & 1.0800 & 0.0227 & 1.7016 & 1.65E-3& 0.189 & 7.9E-3&1.6139&3.5E-4&0.1749&3.6E-3 \\
60 & 1.1953 & 0.0300 & 1.6838 & 2.05E-3& 0.2144 & 0.0107&1.5936&3.5E-4&0.1793&4.45E-3 \\
120& 1.4709 & 0.052 & 1.6500 & 3.1E-3& 0.2681 & 0.019&1.5568&3.5E-4&0.1820&6.5E-3 \\
200& 1.6698 & 0.0718 & 1.6307 & 3.85E-3& 0.2996 & 0.026&1.5387&3E-4&0.1791&7.85E-3 \\
300& 1.8252 & 0.089 & 1.6187 & 4.3E-3& 0.3198 & 0.0316&1.5283&2E-4&0.1748&8.8E-3 \\

\hline\noalign{\smallskip}
\end{tabular}
\end{minipage}
\end{table}
\begin{table}
\centering \caption{The charm structure function determined based
on the reduced charm cross section data that accompanied with
errors. }\label{table:table1}
\begin{minipage}{\linewidth}
\renewcommand{\thefootnote}{\thempfootnote}
\centering
\begin{tabular}{|l||c|c||c|c||c|c||c|c|} \hline\noalign{\smallskip} $Q^{2}(GeV^{2})$ & $ x$ &
$ y$ & $ \widetilde{\sigma}^{c\overline{c}}$ & $
\delta_{\widetilde{\sigma}^{c\overline{c}}}(\%)$ &
$F_{2}^{c\overline{c}}$(Ref.13) &$
\delta_{F_{2}^{c\overline{c}}}(\%)$ & $F_{2}^{c\overline{c}}$(Our
Results) &
 $ \delta_{F_{2}^{c\overline{c}}}$  \\
\hline\noalign{\smallskip}
8.5& 0.00050 & 0.167 & 0.176 & 14.8 & 0.176 & 1.0 &0.1763 &14.8    \\
8.5& 0.00032 & 0.262 & 0.186 & 15.5 & 0.187 & 1.0 &0.1869 &15.6    \\
12 & 0.00130 & 0.091 & 0.150 & 18.7& 0.150 & 1.0 &0.1501 &18.7  \\
12 & 0.00080 & 0.148 & 0.177 & 15.9& 0.177 & 1.1 &0.1773 &15.9  \\
12 & 0.00050 & 0.236 & 0.240 & 11.2& 0.242 & 1.0 &0.2441 &11.4  \\
12 & 0.00032 & 0.369 & 0.273 & 13.8& 0.277 & 1.1 &0.2764 &14.0  \\
20 & 0.00200 & 0.098 & 0.187 & 12.7 & 0.188 & 1.1 &0.1871&12.7 \\
20 & 0.00130 & 0.151 & 0.219 & 11.9 & 0.219 & 1.1 &0.2194&11.9 \\
20 & 0.00080 & 0.246 & 0.274 & 10.2 & 0.276 & 1.0 &0.2756&10.3 \\
20 & 0.00050 & 0.394 & 0.281 & 13.8 & 0.287 & 1.1 &0.2859&14.0 \\
35 & 0.00320 & 0.108 & 0.200 & 12.7& 0.200 & 1.1&0.2002&12.7\\
35 & 0.00200 & 0.172 & 0.220 & 11.8& 0.220 & 1.0&0.2206&11.8 \\
35 & 0.00130 & 0.265 & 0.295 & 9.70& 0.297 & 1.0&0.2973&9.8 \\
35 & 0.00080 & 0.431 & 0.349 & 12.7& 0.360 & 1.1&0.3575&13.0 \\
60 & 0.00500 & 0.118 & 0.198 & 10.8& 0.199 & 1.1&0.1983&10.8 \\
60 & 0.00320 & 0.185 & 0.263 & 8.40 & 0.264 & 1.0&0.2640&8.5 \\
60 & 0.00200 & 0.295 & 0.335 & 8.80 & 0.339 & 1.0&0.3385&8.9 \\
60 & 0.00130 & 0.454 & 0.296 & 15.1& 0.307 & 1.0&0.3047&15.6 \\
120& 0.01300 & 0.091 & 0.133 & 14.1& 0.133 & 1.2&0.1331&14.1 \\
120& 0.00500 & 0.236 & 0.218 & 11.1& 0.220 & 1.1&0.2194&11.2 \\
120& 0.00200 & 0.591 & 0.351 & 12.8& 0.375 & 2.9&0.3712&13.6 \\
200& 0.01300 & 0.151 & 0.161 & 11.9& 0.160 & 2.7&0.1604&11.9 \\
200& 0.00500 & 0.394 & 0.237 & 13.5& 0.243 & 2.9&0.2419&13.8 \\
300& 0.02000 & 0.148 & 0.117 & 18.5& 0.117 & 2.9&0.1173&18.5 \\
300& 0.00800 & 0.369 & 0.273 & 12.7& 0.278 & 2.9&0.2777&12.9 \\
\hline\noalign{\smallskip}
\end{tabular}
\end{minipage}
\end{table}
\begin{figure}
\includegraphics{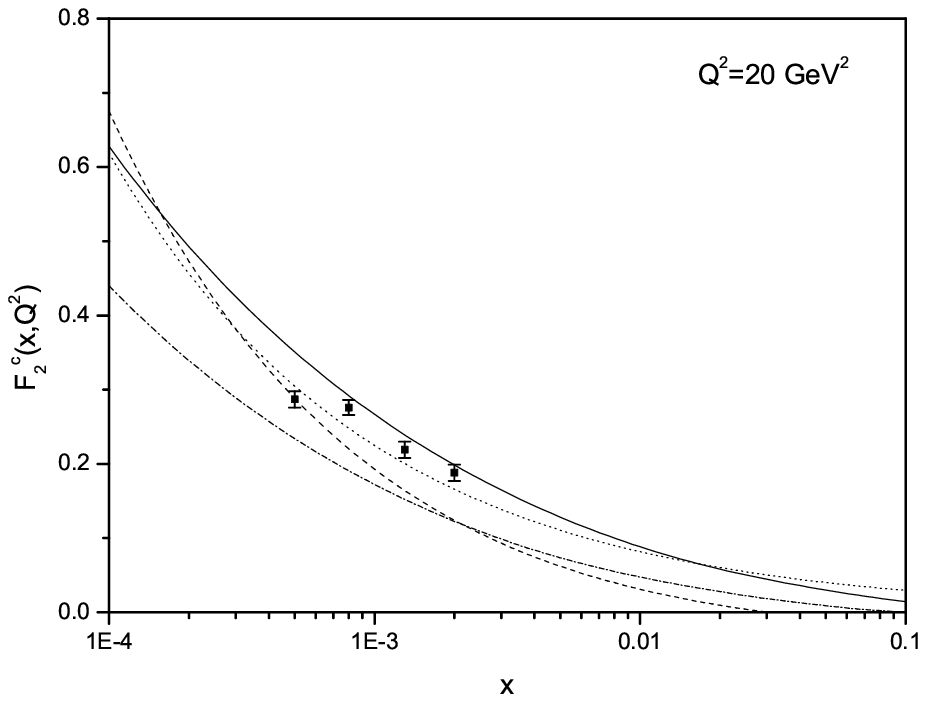}
\caption{ }\label{Fig3a}
\end{figure}
\begin{figure}
\includegraphics{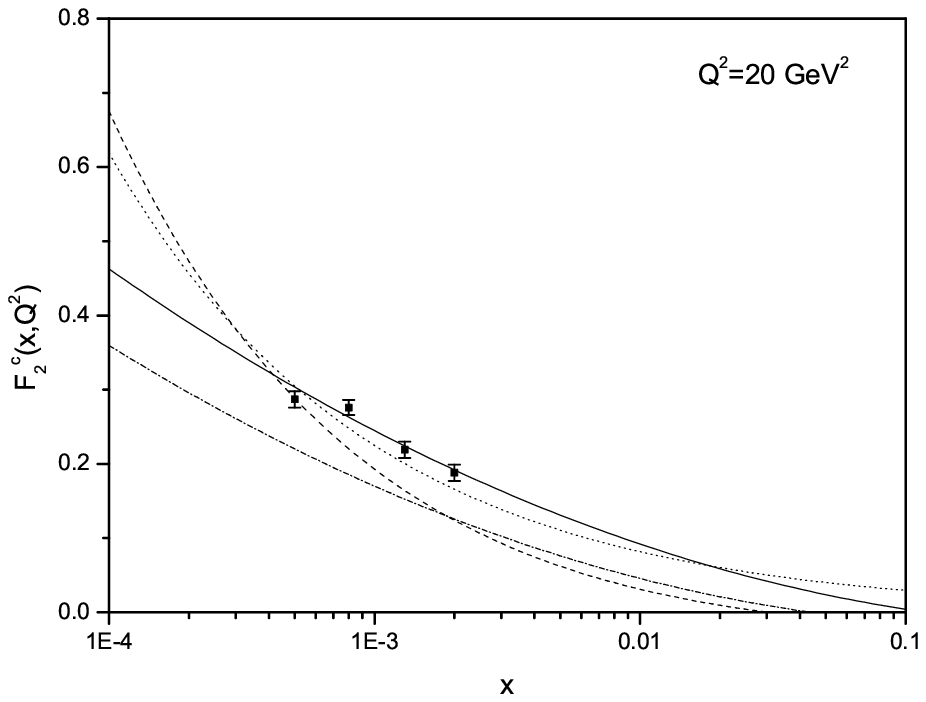}
\caption{  }\label{Fig3b}
\end{figure}
\begin{figure}
\includegraphics{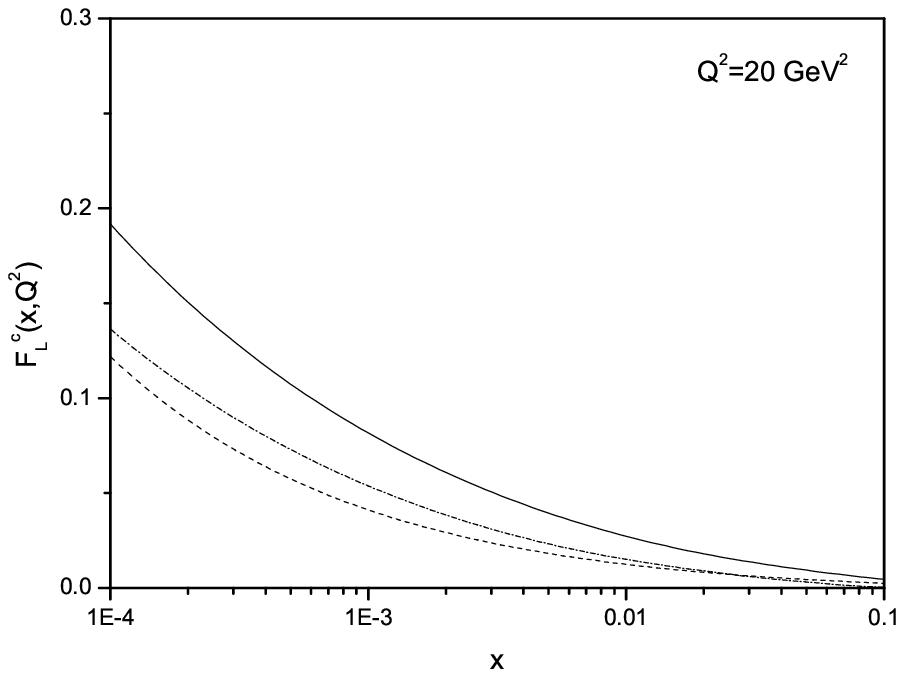}
\caption{ }\label{Fig4a}
\end{figure}
\begin{figure}
\includegraphics{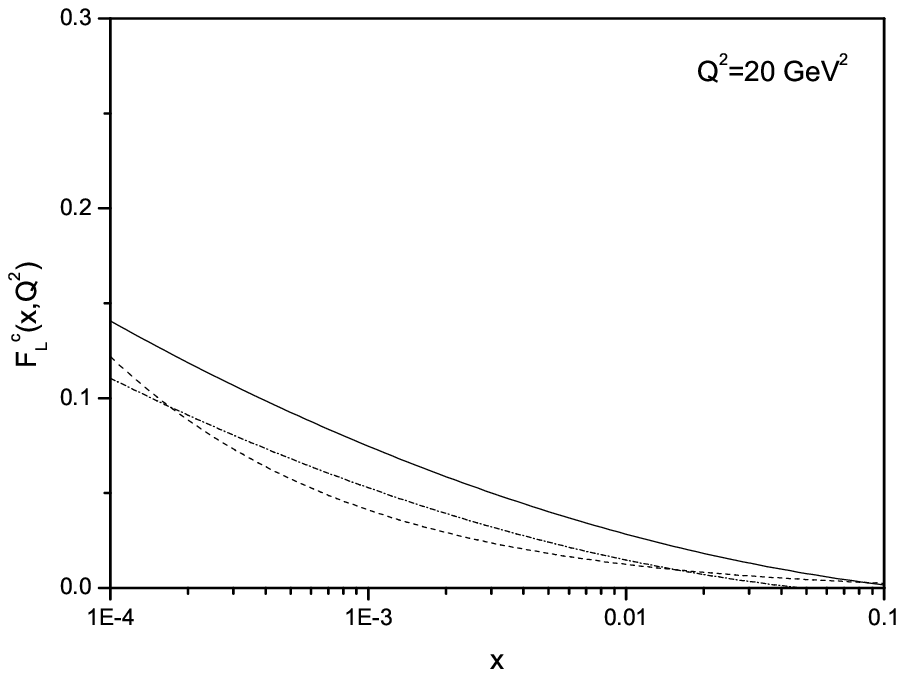}
\caption{  }\label{Fig4b}
\end{figure}
\end{document}